\documentclass[10pt]{article}

\usepackage{amsmath}

\usepackage{array}

\usepackage{appendix}

\usepackage{tocloft}                   % Adds Part structure to table of contents

\usepackage{graphicx}

\usepackage{amsfonts}

\usepackage{amssymb}

\usepackage{mathrsfs}

\usepackage{yfonts}

\usepackage{euscript}

\usepackage{centernot}                 % Permits me to trivially make a centred not version of any object

\usepackage{ifsym}                     % Gives a nicer sharp

\usepackage{lmodern}                   % Gives bold typewriter fonts

\usepackage{upgreek}

\usepackage{mathtools}

\usepackage{color}

\usepackage{slantsc}
\usepackage{calligra}

\usepackage{bbold}          % Gives mathbb acting on lower case and, hopefully, numbers...

\usepackage[T1]{fontenc}

\usepackage{epsf}

\usepackage{latexsym}

\usepackage{tipa}

\usepackage{makeidx}

\makeindex

\def\b{\begin{equation}}

\def\e{\begin{equation}}

\def\be{\begin{equation}}              % Longer older ones kept for rext import compatibility.

\def\ee{\end{equation}}

\def\beq{\begin{equation}}

\def\eeq{\end{equation}}

\def\bea{\begin{eqnarray}}

\def\eea{\end{eqnarray}}

\def\half{\mbox{$\frac{1}{2}$}}

\def\m{\mbox{ }}

\def\mma {\m , \m \m }

\def\!{\hspace{-1.6667em}}

\def\n{\noindent}

\def\u{\underline}

\def\w{\widetilde}

\def\slTheta{\mathit{\Theta}}                     % triangleland azimuth variable. A continuous deformation.

\def\slPhi{\mathit{\Phi}}                         % relative angles in general (suffixed) and relative angle in triangleland in particular (unsuffixed). Also a SIC variable.

\def\ip{i^{\prime}}

\def\jp{j^{\prime}}

\def\ipp{i^{\prime\prime}}

\def\jpp{j^{\prime\prime}}

\def\bic{\mbox{\boldmath$c$}}

\def\bigg{\mbox{\boldmath$g$}}

\def\sbig{\mbox{\scriptsize\boldmath$g$}}

\def\sbic{\mbox{\scriptsize\boldmath$c$}}

\def\bil{\mbox{\boldmath$l$}}

\def\bim{\mbox{\boldmath$m$}}

\def\bip{\mbox{\boldmath$p$}}

\def\biM{\mbox{\boldmath$M$}}

\def\biP{\mbox{\boldmath$P$}}

\def\biQ{\mbox{\boldmath$Q$}}

\def\biS{\mbox{\boldmath$S$}}

\def\sbiM{\mbox{\scriptsize\boldmath$M$}}

\def\sbiN{\mbox{\scriptsize\boldmath$N$}}

\def\sbiQ{\mbox{\scriptsize\boldmath$Q$}}

\def\bbeta{\mbox{\boldmath$\beta$}}

\def\bipi{\mbox{\boldmath$\pi$}}

\def\birho{\mbox{\boldmath$\rho$}}

\def\brho{\birho}                                   % Mass-weighted relative Jacobi configuration space vector

\def\mF{\mbox{F}}

\def\mG{\mbox{G}}

\def\mI{\mbox{I}}                        % Isosceles triangle/ meridian(s) in triangleland

\def\mb{\mbox{b}}

\def\mh{\mbox{h}}

\def\ms{\mbox{s}}

\def\muu{\mbox{u}}

\def\uB{\u{\mbox{B}}}

\def\urho{{\u{\rho}}}

\def\scc{\mbox{\scriptsize c}}

\def\sd{\mbox{\scriptsize d}}

\def\se{\mbox{\scriptsize e}}

\def\sg{\mbox{\scriptsize g}} 

\def\sh{\mbox{\scriptsize h}} 

\def\si{\mbox{\scriptsize i}}

\def\sj{\mbox{\scriptsize j}}

\def\sll{\mbox{\scriptsize l}}  % NB EXCEPTIONAL DEF as \sl is reserved for slant.

\def\sm{\mbox{\scriptsize m}}

\def\sn{\mbox{\scriptsize n}} 

\def\so{\mbox{\scriptsize o}} 

\def\sp{\mbox{\scriptsize p}}

\def\sr{\mbox{\scriptsize r}}

\def\sss{\mbox{\scriptsize s}}  %TO AVOID ARXIV changing \ss to German double s.

\def\st{\mbox{\scriptsize t}}

\def\su{\mbox{\scriptsize u}}

\def\sw{\mbox{\scriptsize w}}

\def\sA{\mbox{\scriptsize A}} 

\def\sB{\mbox{\scriptsize B}}

\def\sC{\mbox{\scriptsize C}}

\def\sI{\mbox{\scriptsize I}}

\def\sJ{\mbox{\scriptsize J}}

\def\sM{\mbox{\scriptsize M}} 

\def\sN{\mbox{\scriptsize N}}

\def\sP{\mbox{\scriptsize P}}

\def\sR{\mbox{\scriptsize R}}

\def\sS{\mbox{\scriptsize S}}

\def\sU{\mbox{\scriptsize U}}

\def\sfa{\mbox{\sffamily{\scriptsize a}}}     % For 1 index less than $\sfA$ in splitting out a scalar time.

\def\sbg{\mbox{{\bf \scriptsize g}}}

\def\sbg{\mbox{{\bf \scriptsize\sffamily g}}}

\def\sbM{\mbox{{\bf \scriptsize M}}}

\def\sbcL{\mbox{\boldmath \scriptsize ${\cal L}$}}

\def\sbcS{\mbox{\boldmath \scriptsize ${\cal S}$}}

\def\usF{\u{\mbox{\scriptsize F}}}

\def\te{\mbox{\tiny e}}

\def\ti{\mbox{\tiny i}}

\def\tm{\mbox{\tiny m}}

\def\tn{\mbox{\tiny n}}

\def\tB{\mbox{\tiny B}}

\def\tJ{\mbox{\tiny J}}

\def\tM{\mbox{\tiny M}}

\def\tS{\mbox{\tiny S}}

\def\tU{\mbox{\tiny U}}

\def\cr{\mbox{\scriptsize{\bf $\m  \times \m $}}}

\def\sumi2{\sum\mbox{}_{\mbox{}_{\mbox{\scriptsize $i$=1}}}^2}

\def\sumi3{\sum\mbox{}_{\mbox{}_{\mbox{\scriptsize $i$=1}}}^3}

\def\sumin{\sum\mbox{}_{\mbox{}_{\mbox{\scriptsize $i$=1}}}^{n}}

\def\sumAn{\sum\mbox{}_{\mbox{}_{\mbox{\scriptsize $A$=1}}}^{n}}

\def\sumABcycles3{\sum\mbox{}_{\mbox{}_{\mbox{\scriptsize cycles $A,B$=1}}}^{3}}

\def\sumCDcycles3{\sum\mbox{}_{\mbox{}_{\mbox{\scriptsize cycles $C,D$=1}}}^{3}}

\def\sumj3{\sum\mbox{}_{\mbox{}_{\mbox{\scriptsize $j$=1}}}^3}

\def\sumk3{\sum\mbox{}_{\mbox{}_{\mbox{\scriptsize $k$=1}}}^3}

\def\sumfand{\sum\mbox{}_{\mbox{}_{\mbox{\scriptsize $\Gamma = 1$}}}^{nd - 1}}                 % preshape space version

\def\prodiA1{\prod\mbox{}_{\mbox{}_{\mbox{\scriptsize $i$=1}}}^{A - 1}}

\def\d{\textrm{d}}                                                  % ordinary derivative

\def\pa{\partial}                                                   % partial derivative

\def\Circ{\mbox{\Large$\circ$}}                                     % Big dot             for hanging things on

\def\Last{\mbox{\Large$\ast$}}                                      % big six-point  star for hanging things on

\def\es{\m = \m}

\def\:={\m := \m}

\def\=:{\m =: \m}

\def\Abs{\mbox{\Large $\mathfrak{a}$}\mb\ms}                         % Absolute space.

\def\FrT{\mathfrak{T}}                                         % A time-line. 

\def\lFrg{\mbox{\Large$\mathfrak{g}$}}                         % Irrelevant group, Lie group.
\def\nFrg{\mbox{\large$\mathfrak{g}$}}                         % Ditto, for use in footnotes and captions.  
\def\LFrg{\mbox{\LARGE $\mathfrak{g}$}}                        % Titlesize

\def\Frg{\mbox{\normalsize $\mathfrak{g}$}}                    % Lie algebra

\def\FrT{\mbox{\boldmath$\mathfrak{T}$}}                       % Used for tangent space and then cotangent space is $^*$ of this. 
\def\Hilb{\mbox{{\boldmath$\mathfrak{H}$}ilb}}                 % Hilbert space
\def\lt{\mbox{\Large $t$}}                                 % ditto: finite

\def\scC{\mbox{\scriptsize ${\cal C}$}}                    % general constraint, regardless of linearity or not in the momenta

\def\scD{\mbox{\scriptsize ${\cal D}$}}                    % zero total dilational momentum constraint

\def\scE{\mbox{\scriptsize ${\cal E}$}}                    % mechanical energy constraint

\def\scH{\mbox{\scriptsize ${\cal H}$}}                    % Hamiltonian constraint of GR.

\def\scL{\mbox{\scriptsize ${\cal L}$}}                    % zero total angular momentum constraint

\def\bLin{\sbcL\mbox{\bf in}} 

\def\Chronos{\scC\mbox{hronos}}                            % Chronos constraint.

\def\bShuffle{\sbcS\mbox{\bf huffle}} 

\def\FrQ{\mbox{\Large $\mathfrak{q}$}}                               % Configuration space
\def\bFrL{\mbox{\boldmath$\mathfrak{L}$}}                            % The space of light degrees of freedom

\def\Phase{\mbox{{\boldmath$\mathfrak{P}$}hase}}                     % Phase space.

\def\bFrR{\mbox{\boldmath$\mathfrak{R}$}}                            % First letter of RigPhase, also used for Riem etc.  Is also, by itself, a ring.
\def\Rig-Phase{\bFrR\mbox{ig-}\Phase}                                % Rigged Phase Space
                                                                													   
\def\bFrR{\mbox{\boldmath$\mathfrak{R}$}}                            % Used in regularity structure symbol
\def\bFrR{\mbox{\boldmath$\mathfrak{R}$}}                            % Used in incipient regularity structure symbol
\def\1mat{\u{\u{1}}}                                                 % unit-entry matrix

\def\Positive-Modespace{\mbox{{\boldmath$\mathfrak{M}$}odespace$^+$}}% Positive modespace

\def\POSITIVE-MODESPACE{\mbox{{\boldmath$\mathfrak{M}$}ODESPACE$^+$}}% Positive modespace alongside scalar field matter inhomogeneous modes.

\def\lE{\mbox{\Large E}}

\def\Kin-Hilb{\mbox{{\boldmath$\mathfrak{K}$}in-\Hilb}}                     % Dynamical Hilbert space 

\def\Mid-Hilb{\mbox{{\boldmath$\mathfrak{M}$}id-\Hilb}}                     % Dynamical Hilbert space 

\def\Dyn-Hilb{\mbox{{\boldmath$\mathfrak{D}$}yn-\Hilb}}                     % Dynamical Hilbert space 

\def\5Star{\mbox{\Large$\star$}}              % Rectified time derviative actually used

\begin{document}

\begin{center}

\Huge{\bf A LOCAL RESOLUTION OF}

\vspace{.1in}

\normalsize

\Huge{\bf THE PROBLEM OF TIME}

\vspace{.15in}

\large{\bf V. Combining Temporal and Configurational Relationalism for Finite Theories}

\vspace{.1in}

\normalsize

{\bf E.  Anderson} 

\vspace{.15in}

{\large \it based on calculations done at Peterhouse, Cambridge} 

\end{center}

\begin{abstract}

As we have known comprehensively since the early 1990's works of Isham and Kucha\v{r}, The Problem of Time mostly concerns interferences between its many facets.  
Having introduced the local facets in Articles I to IV, we now show how Article I's approach to Temporal Relationalism 
                                                   can be combined with Article II's to Configurational Relationalism. 
This requires reformulating some of the Principles of Dynamics to be Temporal Relationalism implementing, 
a strategy which we follow through in each subsequent Article. 
All in all, around half of the Principles of Dynamics needs to be rewritten to solve this noted, 50-year-old and hitherto unresolved foundational problem. 
This stands as sufficient reason to render new resultant Principles of Dynamics -- `TRiPoD' -- a significant and worthwhile development of the Principles of Dynamics. 
This amounts to taking Jacobi's action principle more seriously than Jacobi himself did, or any authors in between: 
to constitute a new starting point for the entirety of the Principles of Dynamics is to be reworked.  
This can moreover be viewed as a mild recategorization necessitated by the Problem of Time.  
While mathematically simple to carry out, 
it requires quite a lot of conceptual developments, by which it is both prudent and useful exposition to present this first for finite rather than 
Field Theoretic examples in the current Article. 
Article VI then extends this approach to Field Theory and GR.  

\end{abstract}

%=========================================================================================================================================================================================
%=========================================================================================================================================================================================
\section{Introduction}
%=========================================================================================================================================================================================
%=========================================================================================================================================================================================

In this fifth Article on the Problem of Time \cite{Battelle, DeWitt67, Dirac, K81, K91, K92, I93, K99, APoT, FileR, APoT2, AObs, APoT3, A-Lett, ABook, A-CBI, I, II, III, IV}, 
we combine Article I's Temporal Relationalism \cite{L, M, BB82, B94I, FileR} and Article II's Configurational Relationalism \cite{L, M, BB82, FileR} for Finite Theories.  
As we have known comprehensively since Isham and Kucha\v{r}'s works \cite{K92, I93}, The Problem of Time mostly concerns interferences between its many facets, 
of which the current Article provides a first classical-level instalment 

\m 

\n This requires reformulating some of the Principles of Dynamics (PoD) to be Temporal Relationalism implementing: {\it TRiPoD} 
(see also \cite{FEPI, FileR, ABook, MBook}), a strategy which we follow through in each subsequent Article. 
All in all, we shall see over Articles V to XIII that around half of the Principles of Dynamics needs to be rewritten to solve this noted, 
50-year-old and hitherto unresolved foundational problem. 
This stands as sufficient reason to render new resultant Principles of Dynamics -- `TRiPoD' -- 
a significant and worthwhile development of the Principles of Dynamics. 
This amounts to taking Jacobi's action principle more seriously than Jacobi himself did, or any authors in between, 
i.e.\ as a new starting point for the entirety of the Principles of Dynamics to be worked out from.  
This can moreover be viewed as a mild recategorization necessitated by the Problem of Time.  
Whle mathematically simple to carry out, 
it requires quite a lot of conceptual developments, by which it is both prudent and useful exposition to present this first for finite rather than 
Field Theoretic examples in the current Article. 
Article VI subsequently extends this approach to Field Theory and GR.  

\m 

\n In outline, Sec 2 formulates TRi as a homothetic Tensor Calculus, whereas Sec 3 presents enough of TRiPoD for unification with Configurational Relationalism.
The combination of Temporal and Configurational Relationalism is effectuated in Sec 4. 
Relational Particle Mechanics (RPM) examples are provided in Sec 5, including use of Article II's `back-cloth' 
and introducing the Hopf--Dragt coordinates for scaled triangleland. 
Sec 6 finally provides the heavy--slow to light--fast split, including classically foreshadowing of some of Semiclassical Quantum Cosmology's issues.

%===================================================================================================================================================================================
%===================================================================================================================================================================================
\section{TRi is a homothetic tensor calculus}
%===================================================================================================================================================================================
%===================================================================================================================================================================================

As we saw in Article I, the more advanced implementation of TRi 
-- the geometrical dual of Manifestly Parametrization Irrelevant implementation, rather than the Manifestly Reparametrization Invariant one -- 
involves configuration-change variables $(\biQ, \d \biQ)$.  

\m 

\n{\bf Structure 1} We now view $\d \biQ$ as a {\it change covector} \cite{FileR}, according it `change weight' 1. 

\m 

\n We use `change weights' to moreover keep track of which standard Principles of Dynamics entities we replace with TRi versions in forming TRiPoD.   

\m 
 
\n These `change weights' can furthermore be identified as a $\mathbb{Z}$-valued version of the homothetic weights;
these are a subcase of the more widely known \cite{York73, AMP} conformal weights. 
See Article XIV for an outline mathematical description of homothetic tensors.

\m 

\n{\bf Structure 2} `Around half' of the Principles of Dynamics entities we make use of are TRi scalars; 
this class of entities is moreover not only large but containing many conceptually significant entities.

\m 

\n{\bf Tricolor Notation} for the progression from the standard PoD to TRiPoD.
We place TRi scalars in the centre of our pictures of the evolution from standard PoD to TRiPoD, on a colourless background. 
We place objects that are not TRi to the left on a blue background, 
and the objects that replace them as reflection images to the right on a red background.

\m 

\n `Tricolor notation' is named after the French flag, which forms this pattern. 
The left, middle and right columns of objects are moreover the TRiPoD's three legs, 
bearing in mind that the middle leg is kept, whereas the left leg is to be replaced with the right one.  
Such figures permit a quick tally of what proportion of the standard PoD is already TRi and what pat needs replacing with TRiPoD, 
such as the `around half' alluded to above. 

\m 

\n This progression into TRiPoD done, it is occasionally further useful to depict just TRiPoD itself 
with further splitting of the red column into classes by change weight.
These are now {\sl centred} on the white TRi-scalar leg; one of Article XIII's summary figures is of this kind.

%===================================================================================================================================================================
\subsection{Some major examples} 
%===================================================================================================================================================================

So far, in Article I we saw that kinetic energy and Lagrangian are to be replaced by kinetic $\d s$ and Jacobi $\d J$ arc elements respectively; 
these are both change covectors.  
The first equality of (I.61) can moreover be read as these two arc elements being simply interrelated by a conformal transformation.  

\m 

\n The {\sl notion} of action ${\cal S}$ itself remains invariant under these reformulations: it is a change {\sl scalar}.
It now moreover has the formula 
\be 
{\cal S} = \int \d J  \m . 
\ee 
\n The definition of generalized momentum clearly needs some rewriting to attain TRi form, along the chain of reformulations  
\be 
\biP  \:=  \frac{\pa L}{\pa {\biQ}^{\prime}}                  \m \mbox{ to } \m 
\biP  \:=  \frac{\pa L_{\lambda}}{\pa \dot{\biQ}}             \m \mbox{ to } \m  
\biP  \:=  \frac{\pa \, \d J }{\pa \, \d \biQ}                \m , 
\label{Mom-Form}
\ee 
where dot is derivative with respect to $t^{\sN\se\sw\st\so\sn}$, and dash is derivative with respect to $\lambda$.
The first two are standard, whereas the third is equivalent to the second by the `cancellation of the dots' Lemma.  
It is the last of these expressions that involves changes, though, being a ratio of such, momentum is itself a change scalar.  

\m 

\n So, whereas TRi places a change weight upon each use of $\FrT(\FrQ)$, it leaves $\FrT^*(\FrQ)$ invariant.
This matches up with Hamiltonian variables being already-TRi,  
and Article VII extends this to phase space being already TRi, by showing that the Poisson bracket is already-TRi. 
This begins to indicate the build-up of elsewise particularly significant structures that so happen also to be already-TRi 
and thus carry over unmodified in forming TRiPoD.  

\m 

\n The forms of $\scE$ and $\scH$ are immediately recognizable as elsewise well-known equations because $\Chronos$ is a change scalar.

%=========================================================================================================================================================================================
%=========================================================================================================================================================================================
\section{Further details of TRiPoD}
%=========================================================================================================================================================================================
%=========================================================================================================================================================================================

\n Article I also provided TRi's Jacobi--Mach equations of motion.  These moreover admit three simplified cases, 
in parallel to Sec I.3.2's three for the standard PoD. 

\m

\n 1) {\it Lagrange multiplier coordinates} $\bim \, \subseteq \, \biQ$ are such that $\d J$ is independent of $\d \bim$, 
$$
\frac{\pa \, \d J}{\pa \, \d \bim}  \es  0  \m .
$$
The corresponding Jacobi--Mach equation is 
\be
\frac{\pa \, \d J}{\pa \bim}        \es  0  \m .
\label{lmel-2}
\ee
\n 2) {\it Cyclic coordinates} $\bic \, \subseteq \, \biQ$  are such that $\d J$ is independent of $\bic$,  
\be 
\frac{\pa \, \d J}{\pa \bic}  \es 0         \m ,
\ee 
while still featuring $\d \bic$: the corresponding {\it cyclic differential}.\footnote{To avoid confusion, 
%OOOOOOOOOOOOOOOOOOOOOOOOOOOOOOOOOOOOOOOOOOOOOOOOOOOOOOOOOOOOOOOOOOOOOOOOOOOOOOOOOOOOOOOOOOOOOOOOOOOOOOOOOOOOOOOOOOOOOOOOOOOOOOOOOOOOOOOOOOOOOOOOOOOOOOOOOOOOOOOOOOOOOOOOOOOOOOOOOOOOOOOOO
`cyclic' in `cyclic differential' just means the same as `cyclic' in cyclic velocity. 
This is as opposed to it carrying any Algebraic Topology connotations, such as `exact differential' or `cycles' 
-- which in e.g.\ de Rham's case are tied to differentials.\label{Cyclic}} 
%OOOOOOOOOOOOOOOOOOOOOOOOOOOOOOOOOOOOOOOOOOOOOOOOOOOOOOOOOOOOOOOOOOOOOOOOOOOOOOOOOOOOOOOOOOOOOOOOOOOOOOOOOOOOOOOOOOOOOOOOOOOOOOOOOOOOOOOOOOOOOOOOOOOOOOOOOOOOOOOOOOOOOOOOOOOOOOOOOOOOOOOOO
%
The corresponding Jacobi--Mach equation is 
\be
\frac{\pa \, \d J}{\pa \, \d \bic}  \es  \mbox{\bf const}                    \m .
\label{cyclic-vel-2}
\ee
3) {\it The energy integral type simplification}.  
$\d J$ is independent of what was previously regarded as `the independent variable $t$', whereby one Jacobi--Mach equation may be supplanted by the first integral
\be
\d J -  \frac{\pa \, \d J}{\pa \, \d \biQ} \d \biQ  \es  \mbox{ constant }   \m .
\label{en-int-2}
\ee
Suppose further that the equations corresponding to 1)
$$
0  \es  \frac{\pa \, \d J}{\pa \bim}(\bar{\biQ}, \d \bar{\biQ}, \bim) \m \mbox{ can be solved for  } \m \bim \m .
$$
One can then pass from 
\be 
\d J (\bar{\biQ}, \d \bar{\biQ}, \bim)
\ee 
to a reduced 
\be 
\d J_{\sr\se\sd}(\bar{\biQ}, \d \bar{\biQ})                                   \m :
\ee 
{\it multiplier elimination}.

\m 

\n Configuration--change space and configuration--velocity space are conceptually distinct presentations of the same tangent bundle $\FrT(\FrQ)$.  
Formulation in terms of change $\d \biQ$ can furthermore be viewed as introducing a {\it change covector}.
This is in the sense of inducing `{\it change weights}' to PoD entities, 
analogously to how introducing a conformal factor attaches conformal weights to tensors.
For instance, $\d s$ and $\d J$ are change covectors as well.
On the other hand, ${\cal S}$ is a change scalar, i.e.\ already-TRi.  

\m 

\n TRiPoD's formulation of momentum is  
\be
\biP  \:=  \frac{\pa \, \d J}{\pa \, \d \biQ}                                 \m ,  
\label{TRi-Mom-2}
\ee
which is a change scalar as well.

%=========================================================================================================================================================================================
\subsection{Free end notion of space variation}\label{FENoS}
%=========================================================================================================================================================================================

\n Suppose a formulation's multiplier coordinate $\bim$ is replaced by a cyclic velocity       $\bic$ \cite{ABFO, FEPI} 
                                                                    or a cyclic differential $\d\bic$ \cite{FileR, ARel}. 

\m 

\n The zero right hand side of the multiplier equation is replaced by $f(\mbox{notion of space alone})$ in the corresponding cyclic equation.  

\m 

\n If the quantity being replaced is an entirely physically meaningless auxiliary, however, 
in the cyclic formulation, the meaninglessness of its values at the end notion of space becomes nontrivial. 

\m 

\n I.e.\ {\it free end point} variation -- a type of {\it variation with natural boundary conditions}) \cite{CH, Fox, BrMa, Lanczos} -- 
is the appropriate procedure.\footnote{To be 
%OOOOOOOOOOOOOOOOOOOOOOOOOOOOOOOOOOOOOOOOOOOOOOOOOOOOOOOOOOOOOOOOOOOOOOOOOOOOOOOOOOOOOOOOOOOOOOOOOOOOOOOOOOOOOOOOOOOOOOOOOOOOOOOOOOOOOOOOOOOOOOOOOOOOOOOOOOOOOOOOOOOOOOOOOOOOOOOOOOOOOOOOO
clear, `free end' here refers to free value {\sl at} the end point 
rather than the also quite commonly encountered freedom {\sl of} the end point's location.} 
%OOOOOOOOOOOOOOOOOOOOOOOOOOOOOOOOOOOOOOOOOOOOOOOOOOOOOOOOOOOOOOOOOOOOOOOOOOOOOOOOOOOOOOOOOOOOOOOOOOOOOOOOOOOOOOOOOOOOOOOOOOOOOOOOOOOOOOOOOOOOOOOOOOOOOOOOOOOOOOOOOOOOOOOOOOOOOOOOOOOOOOOOO

\m 

\n Such a variation imposes more conditions than the more usual fixed-end variation does: three conditions per variation, 
\beq
\frac{\pa \, \d J}{\pa \bigg} = \d \bip^g \mma \mbox{alongside }   \m  
\left.
\bip^g 
\right|_{\mbox{\scriptsize end}} = 0                               \m .  
\label{correct}
\eeq
{\bf Case 1)} If the auxiliaries $\bigg$ are multipliers $\bim$, (\ref{correct}) just reduces to
$$
\bip^{g} \es 0  \mma  \frac{\pa J}{\pa \bim}  \es  0
$$ 
and redundant equations.
So in this case, the end point terms automatically vanish by applying the multiplier equation to the first factor of each.  
This holds regardless of whether the multiplier is not auxiliary and thus standardly varied, or auxiliary and thus free end notion of space varied. 
This is because this difference in status merely translates to whether or not the cofactors of the above zero factors are themselves zero.  
The free end notion of space subtlety consequently in no way affects the outcome in the multiplier coordinate case.  
This may account for the above subtlety long remaining unnoticed. 

\m 

\n{\bf Case 2)} Suppose the auxiliaries $\bigg$ are considered to be cyclic coordinates $\bic$.  
Then (\ref{correct}) reduces to 
\beq 
\left.
\bip^g
\right|_{\se\sn\sd-\sp\st}  \es  0 \m 
\label{ckill}
\eeq
alongside
$$
\dot{\bip}^g = 0 \m  \mbox{(or equivalently} \m \d \bip^g = 0) 
$$
\beq
\m \Rightarrow \m \m
\bip^g  =  C  \mbox{\scriptsize , invariant along the curve of notion of space}  \m . 
\label{hex}
\eeq 
$C(\mbox{notion of space})$ is now identified as $0$ at either of the two end notion of space (\ref{ckill}).
Since this is invariant along the curve of notions of space, it is therefore zero everywhere.  

\m 

\n So (\ref{hex}) and the definition of momentum give 
$$
\frac{\pa L}{\pa \dot{\bic}^{g}}  \:=  \bip^g \m  \mbox{ or equivalently }  \m
\frac{\pa \d J}{\pa \d \bic^{g}}  \es  0                                    \m .  
$$
\n{\bf Remark 1} In conclusion, the above free end point working ensures that the cyclic and multiplier formulations of auxiliaries in fact give {\sl the same} variational equation. 
Thus complying with Temporal Relationalism by passing from encoding one's {\sl auxiliaries} as multipliers 
to encoding them as cyclic velocities or differentials is valid without spoiling the familiar and valid physical equations.

\m 

\n{\bf Remark 2} A similar working \cite{FEPI} establishes that passage to the Routhian for an auxiliary formulated in cyclic terms reproduces the outcome of   
                                                         multiplier elimination for that same auxiliary formulated in terms of multipliers.

%=========================================================================================================================================================================================
\subsection{TRi Legendre transformation}
%=========================================================================================================================================================================================

We next introduce Legendre transformations that interconvert changes $\d \biQ$ and momenta $\biP$. 

\m 

\n{\bf Example 1)} {\it Passage to the $d$-Routhian}  
\beq
\d R(\bar{\biQ}, \d \bar{\biQ}, \biP_{\scc}) \:=  \d J(\bar{\biQ}, \d\bar{\biQ}, \d \bic) - \biP^{\scc} \d \bic  \m .
\label{d-Routh}
\eeq
{\it passage to the} {\it dRouthian} furthermore requires being able to 
$$ 
\mbox{solve  } \m \mbox{\bf const}  \es  \frac{\pa \, \d J}{\pa \, \d \bic}(\bar{\biQ}, \d \bar{\biQ}, \d \bic)
\m \mbox{ as equations for the } \m \d \bic \m .
$$
This is followed by substitution into (\ref{d-Routh}).
One application of this is the passage from Euler--Lagrange type actions to the geometrical form of the Jacobi actions, 
now done without ever introducing a parameter; another is Section II.6's reduction procedure.  

\m 

\n{\bf Example 2)} {\it Passage to the $d$-anti-Routhian}, 
\beq
\d A(\bar{\biQ}, \bar{\biP}, \d \bic)  \es  \d J(\bar{\biQ}, \d \bar{\biQ}, \d \bic) - \bar{\biP}^{\scc} \d \bar{\biQ}  \m .
\eeq
A subcase of this plays a significant role in the next Section.

\m 

\n{\bf Example 3)} {\it Passage to the $\d$-Hamiltonian}, 
\beq
\d H(\biQ, \biP)  \es  \biP \d \biQ - \d J(\biQ, \d \biQ)  \m .  
\eeq
The corresponding equations of motion are now $d${\it -Hamilton's equations} 
\beq
\frac{\pa \, \d H}{\pa \biP}  \es    \d \biQ  \mma  
\frac{\pa \, \d H}{\pa \biQ}  \es  - \d \biP  \m . 
\eeq

%===================================================================================================================================================================================
%===================================================================================================================================================================================
\section{Combining Temporal and Configurational Relationalisms}
%===================================================================================================================================================================================
%===================================================================================================================================================================================

%===================================================================================================================================================================================
\subsection{Configurational Relationalism tailored to also fit Temporal Relationalism}
%===================================================================================================================================================================================

\n Let us firstly consider $\lFrg$-Lie-drag-correcting our incipient Jacobi action's 
\be
{\cal S}_{\sJ}^{\sM\sR\sI} = \int \sqrt{2 \, W} \d s(\biQ, \d \biQ) 
\ee  
label-velocities 
\be 
\dot{\biQ} \longrightarrow \dot{\biQ} - \pounds_{\sbig}\biQ
\ee  
\cite{BB82, B03}. 
This however spoils its Manifest Reparametrization Invariance property, 
since the $\d \lambda$ of integration cancels with the $\d/\d\lambda$'s in the quadratic kinetic term, but not with the corrections.  

\m 

\n We get round this first facet interference by using instead label-velocity representations 
\be 
\dot{\biQ} \longrightarrow \Circ_{\dot{\sbic}}\biQ = \dot{\biQ} - \pounds_{\dot{\sbic}} \biQ
\ee 
of the generators \cite{ABFO}. 
This succeeds in preserving Manifest Reparametrization Invariance, 
but shifts the auxiliary $\lFrg$-variables' PoD status from multipliers to cyclic coordinates. 
None the less, {\it free end point value variation} \cite{FEPI, FileR} 
ensures the outcome of varying with respect to the $\lFrg$-variables transcends this reformulation.  
Let us also identify this `best-matched' solution followed by substituting back into the original action as a Routhian Reduction \cite{Lanczos}. 

\m  

\n To work with the geometrical action, moreover, we require \cite{FileR} to Lie-drag-correct with $\lFrg$-auxiliaries that are themselves changes, 
\be 
\d{\u{\biQ}} \m \longrightarrow \m \d_{\sg} \biQ  \:=  \d{\u{q}}^I - \pounds_{\d \sbg}{\u{\biQ}}    \m .  
\label{Lie-Drag-2}
\ee    
Free end point value variation carries over. 
Our procedure now moreover involves not passage to the Routhian but a TRi-variant thereof: passage to the {\it dRouthian} \cite{TRiPoD, AM13}. 
Namely, elimination from the original action of the properly-TRi Jacobi--Mach variables formulation's cyclic differentials 
                                                                                       rather than of cyclic velocities in the usual passage to the Routhian.
Using this final method, we succeed in jointly incorporating Temporal and Configurational Relationalism at the level of the action.  

\m 

\n In greater generality, we need to adjust the $\lFrg$-act $\lFrg$-all Method's $\lFrg$ auxiliaries to be TRi-represented 
to jointly incorporate Temporal and Configurational Relationalism. 

\m 

\n Note finally that we need to {\sl iterate} implementing Temporal and Configurational Relationalisms until consistency is attained.  
This is represented by the loop in Fig 1.b). 

\m 

\n A first reason for this is that finding an explicit $t^{\se\sm}$ 
in the presence of Lie-drag corrections requires prior elimination of the $\lFrg$-auxiliary variables. 
This is clear from the $d\biQ$ in its formula still carrying these corrections unless one substitutes in the best-matched values for these, 
or, alternatively, starts afresh with the reduced action after performing this elimination therein.  
The next subsection's facet's check provides a second reason.

%=================================================================================================================================================================================
\subsection{Preliminary considerations}
%=================================================================================================================================================================================

We now reach our first systematic combination of Problem of Time facets.
Two preliminary considerations are as follows.  

\m

\n Firstly, continuing from Article II's discussion about Configurational Relationalism's postulate i), 
at the level of standard redundant presentations, the mechanical 
\be 
M_{iIjJ} = m_I\delta_{ij}\delta_{IJ}
\ee 
is unsatisfactory through involving absolute space's Euclidean metric: an extraneous spatial structure.  

\m 

\n However, we show in Article VI that the GR counterpart succeeds in meeting this criterion.

\m

\n Secondly, we rephrase Configurational Relationalism ii) to reflect that $\lFrg$ more strictly acts not on $\FrQ$ but on some fibre bundle structure thereover, 
such as $\FrT(\FrQ)$ in the case of Best Matching. 
It additionally applies more generally to further structures based upon $\FrQ$ in the case of the wider range of examples 
covered by the $\lFrg$-act, $\lFrg$-all method.  

\m

\n This Section, moreover, mostly concerns various generalizations and reformulations of the Best Matching implementation we already encountered in Article II.
We now extend this to general $\lFrg$ and then render it TRi-compatible.

%=========================================================================================================================================================================================
\subsection{TRi Best Matching: General $\LFrg$}\label{Q-G-Comp}
%=========================================================================================================================================================================================

\n We next consider the detailed form taken by Best Matching.  

\m 

\n{\bf TRi Best Matching 0)} The incipient bare action can be thought of as a map 
\be 
{\cal S}: \FrT(\FrQ) \rightarrow \mathbb{R}                                                 \m .
\ee
We now interpret the tangent bundle $\FrT(\FrQ)$ as a configuration--change space with a product-type Jacobi action thereupon.   
We next pass to the `arbitrary $\lFrg$ frame corrected' PoD action 
by applying the basic infinitesimal group action to the incipient bare ${\cal S}$, obtaining 
\beq
{\cal S}_{\sR}     \es   \int \d J 
                   \es   \sqrt{2} \int \sqrt{W(\biQ)} \d s  \mma 
\ee 
\be
\d s               \:=   {||\d_{\sg} \biQ||_{\sbiM}}^2
                   \es   {||\dot{\biQ} - \stackrel{\rightarrow}{\lFrg}_{\sbig} \biQ||_{\sbiM}}^2   \m .  
\eeq
\n{\bf Remark 1} This $\lFrg$ is to be suitably compatible with $\FrQ$ in the sense of Sec II.5.2's criteria A) to C).  

\m

\n{\bf Remark 2} A given $(\FrQ, \lFrg)$ pair still constitutes a substantial ambiguity as regards which action ${\cal S}$ to consider thereupon.  

\m 

\n{\bf Remark 3} Relationalism does not highly uniquely control the form that Theoretical Physics is to take.   

\m 

\n{\bf TRi Best Matching 1)} We next extremize over $\lFrg$. 
This produces a constraint equation $\bShuffle$ of the form $\bLin$, which is linear in the momenta and also a change scalar. 

\m 

\n{\bf Remark 4} The theory is then trivial if 
\be 
\mbox{{\it trivial} if } \m c \m > \m  k            \m ,
\ee  
\be 
\mbox{{\it inconsistent} if } \m c = k              \m ,
\ee  
and 
\be 
\mbox{{\it relationally trivial} if } \m c = k - 1  \m .
\ee 
Since independent constraints use up at least 1 degree of freedom each, 
\be 
c \m \geq \m  l + 1
\ee 
is a guaranteed least stringent bound (the 1 arising from $\Chronos$).   
I.e.\ 
\be 
c  =  l + 1 \m \geq \m k - 1 \, \m \Rightarrow \m \, k - l  
   =  \mbox{dim}(\FrQ) - \mbox{dim}(\lFrg)  \m \leq \m  2
\ee 
already serves to invalidate theories.  

\m

\n Article VII shall moreover replace this rather crude bound with a tighter bound based on phase space geometry rather than mere counting.  
This arises from considering brackets relations in addition to the group actions of generators on (some fibre bundle over) $\FrQ$; 
groups are, of course characterized by generators {\sl and} relations.

\m 

\n{\bf TRi Best Matching 2)} The Machian variables $(\biQ, \d \biQ)$ form of this constraint is to be solved for the auxiliary variables $\bigg$ themselves.

\m 

\n The extremizing values solving this are, schematically, 
\be 
\d \bigg = \d \bigg_{\sB\sM}(\biQ, \d\biQ)   \m . 
\label{BM-val}
\ee 
The final extremized action is then 
\be
{\cal S}_{\sB\sM}(\biQ, \d\biQ)  \es  {\cal S}_{\nFrg\mbox{\scriptsize -free}}  
                          \es  {\cal S}_{\sC\sR}(\biQ, \d\biQ, \d \bigg_{\sB\sM})
                          \=:  {\cal S}_{\sC\sR}(\biQ, \d\biQ \mbox{ alone })						       \m , 
\ee 
using the dependence (\ref{BM-val}) in the final step.

\m 

\n{\bf TRi Best Matching 3)} Substitute this solution back into the action; this is an example of Sec I.3.3's multiplier elimination.  
This produces a final $\lFrg$-independent expression that could have been arrived at as a {\sl direct} implementation of Configurational Relationalism ii). 

\m 

\n{\bf Best Matching 4)} We finally elevate this new action to serve as a new starting point.  

\m 

\n{\bf Remark 5} Overall, 
\beq
{\cal S}_{\sR\si}  \:=  {\cal S}_{\nFrg\mbox{\scriptsize -free}} 
              \es  {\cal S}_{\sB\sM}  
              \:=  \lE_{\d \sbic \, \in \,  \nFrg} \big( {\cal S}_{\sC\sR}(\biQ, \d \biQ, \bigg) \d\bic)  \m . 
\eeq
The first symbol means `the Configurationally Relational PoD action' 
(corresponding to the group action of $\lFrg$ on $\FrT(\FrQ)$ being physically irrelevant). 
This is alternatively known by the second symbol's conceptual type: the $\lFrg$-free PoD action. 
The third symbol is its Best matching implementation, and the final right-hand-side has further computational contact. 

\m

\n{\bf Remark 6} Being a TRi resolution of Configurational Relationalism, 
we denote it with the subscript `Ri' standing for (Temporal and Configurational) Relationalism implementing: 
a 2-aspect implementing or 2-facet overcoming structure. 

\m 

\n{\bf Remark 7}  (TRi) Best Matching 1), 2) and 4) can be viewed as searching for a `minimizer', 
so as to establish the minimum `incongruence between' adjacent physical configurations.
`Minimizer', and indeed the `bestness' of Best Matching are moreover modulo the Calculus of Variations 
not guaranteeing that extremization yields either a minimum or a unique answer. 

\m

\n{\bf Remark 8} These steps can also be viewed as a $\FrT(\FrQ)$-level reduction procedure.\footnote{In this Series of Articles, 
%OOOOOOOOOOOOOOOOOOOOOOOOOOOOOOOOOOOOOOOOOOOOOOOOOOOOOOOOOOOOOOOOOOOOOOOOOOOOOOOOOOOOOOOOOOOOOOOOOOOOOOOOOOOOOOOOOOOOOOOOOOOOOOOOOOOOOOOOOOOOOOOOOOOOOOOOOOOOOOOOOOOOOOOOOOOOOOOOOOOOOOOOO
and quite commonly in the Quantum Gravity literature, reduction usually involves taking into account just linear constraints 
rather than the quadratic constraint as well. \label{Red-Nomen}}
%OOOOOOOOOOOOOOOOOOOOOOOOOOOOOOOOOOOOOOOOOOOOOOOOOOOOOOOOOOOOOOOOOOOOOOOOOOOOOOOOOOOOOOOOOOOOOOOOOOOOOOOOOOOOOOOOOOOOOOOOOOOOOOOOOOOOOOOOOOOOOOOOOOOOOOOOOOOOOOOOOOOOOOOOOOOOOOOOOOOOOOOOO

\m 

\n{\bf Remark 9} We see that in practice that (TRi) Best Matching 2) is often an impasse.  

\m

\n{\bf Remark 10} Best Matching 1), 3) and 4) work out the same whether TRi or not, with 1) and 3)'s {\sl mathematical procedures} also unaffected by this.

\m  

\n{\bf Remark 11} Best Matching moreover involves probing with candidate generators of irrelevant motions, 
without yet addressing the relations between the generators, which would complete the characterization of any given group.
The group relations part of Group Theory enters at a later stage, 
though consideration of the constraints that ensue and whether Constraint Closure applies to these.  

\m

\n{\bf Remark 12} The Lie derivative can also be interpreted as a point identification map \cite{Stewart} {\sl between} two adjacent space slices, 
as opposed to a Lie dragging {\sl within} a single such slice.  

\m 

\n The Lie derivative applies due to continuous transformations being implemented; 
moreover all of those under consideration happen to also be differentiable.  
These transformations form some subgroup 
\be 
\lFrg \m \leq \m  Diff(\Abs)
\ee 
corresponding to 
\be 
\Frg \m \leq \m  Diff(\Abs)
\ee 
at the level of Lie algebras. 

\m  

\n{\bf Remark 13} Let us also briefly comment on descriptions \cite{MBook} of Best Matching that emphasize fibre bundles \cite{IBook, Husemoller} 
                                                                                          rather than Lie derivatives. 
Consistency between the two is enhanced by the notion of Lie derivative continuing to make sense within a fibre bundle context (see e.g.\ \cite{Kobayashi}).  

\m  

\n On the one hand, an advantage of the Lie derivative formulation is that it provides a specific form for the infinitesimal group action. 

\m 

\n On the other hand, $\lFrg$-act does always involve this particular group action (for all that Best Matching itself does). 

\m 

\n A disadvantage is that the Lie derivative offers merely a local treatment, whereas fibre bundles can encode further global information. 
The current Series makes very little nontrivial use of fibre bundle concepts, 
as in some spaces that objects belong to may be identified as fibre bundles but specific global consequences of these being fibre bundles are seldom considered.

%=========================================================================================================================================================================================
\subsection{Emergent Jacobi--Barbour--Bertotti time} \label{temJBB}
%=========================================================================================================================================================================================

\n{\bf TRi-Best Matching} 3$^{\prime}$) As a distinct application of TRi Best Matching 2) -- the emergent time expression is now 
\beq
\lt^{\se\sm}_{\sR\si}  \:=  \lt^{\se\sm}_{\nFrg\mbox{\scriptsize -free}}  
                       \:=  \lE^{\prime}_{\sbig \, \in \, \nFrg} \int   \frac{  ||\d_{\sbig}\biQ||_{\sbiM}  }{  \sqrt{2 \, W(\biQ)}  }       \m . 
\label{Kronos}
\eeq
This is also known as Jacobi--Barbour--Bertotti emergent time, 
but our choice of suffices is a rather clearer reminder of this being an emergent time that complies with Configurational Relationalism.  
Also note how the extremization now takes an implicit form.  
I.e.\ in (\ref{Kronos}) a second functional -- the emergent time -- is subject to performing the extremization of a first functional: 
the relational action ${\cal S}_{\sR\si}$.
This means that 
$$
\lE^{\prime}_{\sbig \, \in \,  \nFrg}  \es  \left.\right|_{\d \sbig = \d \sbig_{\tB\tM}(\sbiQ, \d \sbiQ)}                                    \m ,
$$ 
i.e.\ restriction to the Best-Matched value.
Moreover, if one succeeds in carrying out Best Matching as e.g.\ per Sec \ref{red-RPM}, $\d \bigg$ is replaced by an extremal expression in terms of $\biQ$ and $\d \biQ$ alone.  
By this, both aspects of this complication are washed away and one has an expression for $\lt^{\se\sm}_{\nFrg\mbox{\scriptsize -free}}$ 
                                              paralleling that of the previous Section's $\lt^{\se\sm}$, albeit now in terms of the {\sl reduced} $\FrQ$'s geometry.

%=========================================================================================================================================================================================
%=========================================================================================================================================================================================
\section{Examples}
%=========================================================================================================================================================================================
%=========================================================================================================================================================================================

%=========================================================================================================================================================================================
\subsection{Example 1) Euclidean RPM}\label{Intro-RPM-Ex} 
%=========================================================================================================================================================================================

{\bf Structure 1} In TRi form, the action for this is 
\be 
S  \es  \sqrt{2}\int\sqrt{ E - V(\u{\rho}^A\cdot\u{\rho}^E \mbox{ alone)}} \, \d s             \mma
\ee 
for  
\be 
\d s                                     =  ||\d_{{\u{B}}}\brho||                                      \mma               
\d_{\u{B}} \, \u{\rho}^{A}  :=  \d\u{\rho}^A - \d{\u{B}} \cr \u{\rho}^{A}  \m .  
\label{wasT}
\ee
\n{\bf Structure 2} The conjugate momenta are given by the momentum--change relation
\be 
\bipi \es  \frac{\sqrt{2 \, W} \, \d_{\u{B}}\brho}{||\d_{\u{B}}\brho||}                                        \m .
\ee   
\n{\bf Structure 3} These obey as a primary constraint the quadratic constraint 
\be
\scE  \:=  \half ||\bipi||^2 + V(\brho) 
      \es  E                                                                                                   \m .
\label{HamT}
\ee 
\n{\bf Structure 4} Also free end point variation with respect to $\u{B}$ gives as a secondary constraint the linear zero total angular momentum constraint 
\be
\underline\scL  \:=  \sumin \u{\rho}^{A} \cr \u{\pi}_{A}  
                \es                               0                                                            \m .  
\label{ZAM-2}
\ee
\n{\bf Structure 5} The Jacobi--Mach equations of motion are 
\beq
\frac{\sqrt{2 \, W} \, \d \bipi}{||\d_{{\u{B}}} \birho||}  \es  - \frac{\pa V}{\pa\brho}               \m .
\eeq  
\n{\bf Structure 6} The emergent time here takes the form    
\beq
\lt^{\se\sm}_{\sR\si} \es  \mbox{\large E}^{\prime}_{\u{B} \, \in \,  Rot(d)}  \int  \frac{||\d_{\u{B}}\birho||}{\sqrt{2 \, W(\birho)} }   \m .   
\label{Anima}
\eeq
{\bf Structure 7} In terms of this, the momentum-change relation and equations of motion take the simplified forms 
\beq
\bipi  =  \mathbb{I} \Last \brho \mma \Last \bipi 
      \es  - \frac{\pa V}{\pa\brho}                                                                           \m \mbox{ for } \m 
\eeq
\beq
\Last  \:=  \frac{\d }{\d t^{\se\sm}_{\sR\si}} 
       \:=  \mbox{\large E}^{\prime}_{\u{B} \, \in \,  Rot(d)}  \sqrt{2 \, W} \frac{\d}{||\d_{\u{B}}\brho||}  \m .  
\eeq 
{\bf Remark 1} the 1- and 2-$d$ cases are included within the 3-$d$ form of the auxiliaries.    
There is no $B$ for $d$ = 1, and but a scalar ${B}$ for $d$ = 2. 
All these cases are encoded by 
\be 
\u{\rho}^A -  \u{B} \cr \u{\rho}^A 
\ee 
if one allows for 
\be 
\u{B} = (0, 0, B)
\ee 
in 2-$d$ and 
\be 
\u{B} = 0
\ee 
in 1-$d$ (Exercise!). 

\m 

\n {\bf Remark 2}
Correspondingly, in 1-$d$ there is no $\u{\scL}$ constraint at all, while in 2-$d$ $\u{\scL}$ has just one component that is nontrivially zero: 
\be 
\scL  \es  \sumAn \{\rho^{A1}\pi_{A2} - \rho^{A2} \pi_{A1}\}  
      \es                             0                       \m .
\ee

%=========================================================================================================================================================================================
\subsection{Example 2) similarity RPM}
%=========================================================================================================================================================================================

\n{\bf Structure 1} The geometrical action for this is 
\be 
S  \es  \sqrt{2} \int \sqrt{E - V(\mbox{ratios of } \urho^A\cdot\urho^E \mbox{ alone})} \, \d s \mma 
\ee 
for  
\be 
\d s^2                              \es  \frac{1}{\mI} ||\d_{\u{B},C}\mbox{\brho}||\mbox{}^2              \m \mbox{ and } \m 
\d_{{{\u{B}},C}}\urho^{A}   :=   \d{\urho}^{A} - \d{\uB} \cr \urho^{A} + \d C\urho^A                      \m .
\label{S-SRPM}
\eeq 
{\bf Structure 2} $\d s^2$ is also a ratio since 
\be 
I = ||\brho||^2  \m .  
\ee 
{\bf Structure 3} The conjugate momenta are given by the momentum--change relation 
\be 
\u{\pi}_A  \es  \frac{\sqrt{2 \, W}}{I}\delta_{AB} \, \d_{\u{B},C}\u{\rho}^B              \m .
\ee 
The quadratic energy constraint is 
\beq
\scE  \:=  \frac{I}{2} ||\bipi||^2 + V  
      \es               E                                                                                         \m .
\eeq
{\bf Structure 4} (\ref{ZAM-2})'s $\u{\scL}$ from variation with respect to $\u{B}$ is now accompanied by the {\it zero total dilational momentum constraint}  
\beq
\scD  \:=  \sumAn \m \urho^A \cdot \u{\pi}_A 
      \es                       0 
\label{ZDM}
\eeq
as a secondary constraint from variation with respect to $C$. 

\m 

\n The Jacobi--Mach equations of motion are rather similar to the previous section's.  

\m

\n $t^{\se\sm}$ also parallels the previous section's except that its extremization is now also over $Dil$ of course. 

\m

\n{\bf Remark 2} $\u{\scL}$ and $\scD$ are entirely independent by a standard commutation 
the former acts on shape alone while the latter trivializes the role of scale.

\m 

\n See Fig \ref{BM-Suite} for Best Matching for the Euclidean and similarity cases. 
%
%FFFFFFFFFFFFFFFFFFFFFFFFFFFFFFFFFFFFFFFFFFFFFFFFFFFFFFFFFFFFFFFFFFFFFFFFFFFFFFFFFFFFFFFFFFFFFFFFFFFFFFFFFFFFFFFFFFFFFFFFFFFFFFFFFFFFFFFFFFFFFFFFFFFFFFFFFFFFFFFFFFFFFFFFFFFFFFFFFFFFFFFFF
{            \begin{figure}[!ht]
\centering
\includegraphics[width=0.3\textwidth]{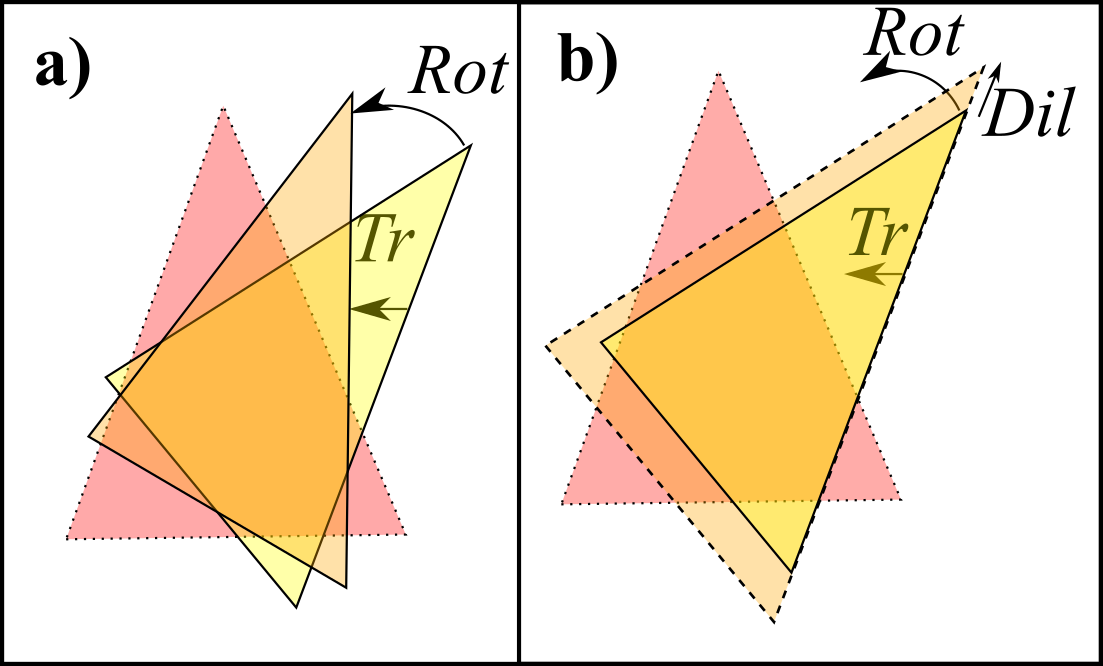}
\caption[Text der im Bilderverzeichnis auftaucht]{        \footnotesize{Two notions of matching shapes 
(keep the red one fixed and perform transformations on the yellow one so as to bring the two into minimum incongruence).

\m 

\n a) Shuffling one triangle relative to the other, subjected to translational and rotational matching, as manifested by solid-object triangles.

\m 

\n b) The same, now including also dilational matching, 
as exhibited e.g.\ by the overhead projector image of a triangle drawn on one slide being moved away from another such.} }
\label{BM-Suite} \end{figure}          }
%FFFFFFFFFFFFFFFFFFFFFFFFFFFFFFFFFFFFFFFFFFFFFFFFFFFFFFFFFFFFFFFFFFFFFFFFFFFFFFFFFFFFFFFFFFFFFFFFFFFFFFFFFFFFFFFFFFFFFFFFFFFFFFFFFFFFFFFFFFFFFFFFFFFFFFFFFFFFFFFFFFFFFFFFFFFFFFFFFFFFFFFFF

\m 

\n Let us end by noting that one can view `Newtonian Mechanics for island universe subsystems within RPM' as a relational route to Newtonian Mechanics.

%=========================================================================================================================================================================================
\subsection{RPM examples of Best Matching solved}\label{red-RPM}
%=========================================================================================================================================================================================

\n Solving Configurational Relationalism for an RPM renders its action $\lFrg$-free and the expression for its 
$t^{\se\sm}_{\nFrg\mbox{\scriptsize -free}}$ explicit. 
As detailed in \cite{FileR}, this can be done in 1- or 2-$d$ for any $N$ and for 3 particles in 3-$d$, in each case with or without scale.
These need particular names as whole-universe models so as to not confuse them with $N$-body problems in their usual subsystem context.
We term these, respectively, {\it N-stop metroland}, {\it N-a-gonland} and {\it triangleland}.  
Triangleland refers to 3 particles in 2-$d$, and {\it quadrilateralland} to 4 particles in 2-$d$. 
For Shape and Scale RPM,   
\beq
S_{\sr\se\sd}     \es  \sqrt{2}\int\sqrt{W}\d s_{\sr\se\sd}       \m \mbox{ for } \m 
\d s_{\sr\se\sd}  :=   \d s_{\sB\sM} = ||\d\biQ||_{\sbiM(\sbiQ)}  \m , 
\eeq
where the $\biM(\biQ)$ are in-general-curved $\FrQ$ metrics. 

\m

\n The above also arises by a $\FrT(\FrQ)$-level classical reduction (\cite{FileR} and Sec II.6 in outline)
and is furthermore the basis for Reduced Quantization (see Article IV).   
Upon having performed the above reduction, the RPM is in a form in which the following absolute structures of the Newtonian and Galilean Paradigms are absent: 
not just $t$ but also velocity $\u{v}$ and flat spatial metric $\delta$.   
Indeed, the reduced configuration space metric is here a function of the configurations themselves rather than involving any further extraneous background entities.  

\m 

\n{\bf Remark 1} RPMs with Newtonian Gravitation potentials have {\sl not} freed themselves from the non-dynamical connection absolute structure of 
                             Newtonian Mechanics with Newtonian Gravitation potential \cite{Ehlers73}.

%===================================================================================================================================================================
\subsection{Direct construction} 
%===================================================================================================================================================================

\n{\bf Remark 1} The idea here is to construct a $\lFrg$-invariant formulation by working directly \cite{FileR} on $\w{\FrQ} = \FrQ/\lFrg$, 
here viewed a priori as {\it relational configuration space} rather than as reduced configuration space.  
This eliminates the need for any arbitrary $\lFrg$-frame variables, nor do any linear constraints arise, 
nor are such to be used as a basis for reduction to pass to a new action. 
One's action is instead now already directly $\lFrg$-invariant i.e.\ on $\w{\FrQ}$ by construction.  
The diagnostic, and price to pay for this, is the ensuing more complicated form taken by the kinetic term or arc element.  

\m

\n{\bf Example 1} Direct construction is another procedure that is practically implementable for lower-$d$ RPMs.  
It involves case starting directly with the shape spaces that Kendall had determined \cite{Kendall84, Kendall} in the pure-shape case,    
or the cones thereover in the scaled case \cite{FileR}. 
In each case, one is associating a mechanics to the geometry which then plays the role of that mechanics' configuration space, 
as per the methodology of Jacobi and Synge \cite{Lanczos} (Sec I.4).
The latter `relational configuration space formulation'  indeed provides a second foundation for RPMs \cite{FORD, FileR}, distinct from that in \cite{BB82, B03}. 
Because the same formulation of the same mechanics arises both ways \cite{FileR} -- on relational configuration space or by reduction -- 
we subsequently refer to it as the {\it r-formulation} (denoted by an r subscript).   

\m

\n Both are analytically tractable in some cases, 
because e.g.\ some 1-$d$ models and some 3-particle models for $d > 1$ have particularly simple configuration spaces, as per Sec II.6.   
In this way, $\lt^{\se\sm}_{\nFrg\mbox{\scriptsize -free}}$ is itself a $\lt^{\se\sm}$ for a more directly formulated $\FrQ$ geometry. 

\m

\n{\bf Remark 2} These identifications of RPM configuration spaces with well-known geometries 
are useful by allowing for very close to standard mathematical treatment at the classical and quantum levels.
This mathematics is moreover physically interpreted in a very different manner from the standard one \cite{FileR}. 
This can be tied to a wider range of still fairly standard mathematical methods so as to do Classical and Quantum Physics.

\m

\n{\bf Remark 3} Triangleland is the simplest model that can concurrently possess scale and nontrivial linear constraints.  
On the other hand, quadrilateralland \cite{QuadI} is more geometrically typical for an $N$-a-gonland than triangleland 
(which benefits from the $\mathbb{CP}^1 = \mathbb{S}^2$ coincidence).  
These models' mathematical simpleness is a triumph, because one can then carry out and check many Problem of Time calculations. 
[These would {\sl not} make sense if done for atoms or molecules, say, due to the absolutist underpinnings in these latter cases.] 
This is just what the study of the Problem of Time needs (at least as regards the facets and strategies nontrivially manifested by RPMs).  

\m

\n{\bf Remark 4} The same geometrical structures arise from Sec \ref{red-RPM}'s Best Matching.

%===================================================================================================================================================================							 
\subsection{Hopf--Dragt coordinates}
%===================================================================================================================================================================

\n{\bf Structure 1} As regards the corresponding relationalspaces, 3-stop metroland's is trivially $\mathbb{R}^2$; indeed $N$-stop metroland's is $\mathbb{R}^{n}$. 
In each case it is entirely clear how to represent an $n$-sphere within $\mathbb{R}^n$.  
The $\u{n}_{A}$ play the role of Cartesian directions.

\m 

\n{\bf Structure 2} What plays this role for 2-$d$ triangleland's $\mathbb{S}^2$ within its relative space $\mathbb{R}^4$ takes more work.    
It turns out that 
\be 
\mathbb{R}^4 \longrightarrow \mathbb{S}^3 \longrightarrow \mathbb{S}^2 \longrightarrow \mathbb{R}^3
\ee 
handles this, where the second step is the `Hopf map', and the other two steps are, more obviously, taking the unit-sphere preshape space and a coning. 
In this manner, the `Hopf--Dragt' quantities \cite{Dragt, Iwai87, LR97} arise:
\beq
dra_x := \mbox{sin}\,\slTheta\,\mbox{cos}\,\slPhi = 2 \, n_1 n_2 \ ,\mbox{cos}\,\slPhi = 2 \{\u{n}_1 \cr \u{n}_2\}_3 \m ,
\label{dragt1}
\eeq
\beq
dra_y := \mbox{sin}\,\slTheta\,\mbox{sin}\,\slPhi = 2 \, n_1 n_2\,\mbox{sin}\,\slPhi = 2 \, \u{n}_1 \cdot \u{n}_2 \m ,
\label{dragt2}
\eeq
\beq
dra_z := \mbox{cos}\,\slTheta = n_2\mbox{}^2 - n_1\mbox{}^2 \m .   
\label{dragt3}
\eeq
These appear as `ubiquitous quantities' \cite{FileR} in studying the kinematics and the relational dynamics of triangle configurations; 
we shall encounter them again in discussing observables for this model arena in Article VIII.  
%
% They are configurational \K observables for the triangleland RPM \cite{AObs}.  

\m 

\n{\bf Structure 3} Using 
\be 
\rho_i =\rho\, n^i \m, 
\ee 
these quantities are also available in scaled form; I denote these by Dra$_i$.

\m

\n{\bf Remark 1} Geometrical interpretation of the Hopf--Dragt quantities is as follows. 

\m 

\n i) $dra_x$ is a quantifier of `anisoscelesness' $aniso$: departure from the underlying clustering's notion of isoscelesness, c.f.\ anisotropy in Sec I.6. 
Specifically, $aniso$ per unit base length in mass-weighted space is the $l_1 - l_2$ indicated in Fig \ref{Tutti-Tri-3}.a):                   
the amount by which the perpendicular to the base fails to bisect it (which it would do were the triangle isosceles).

\m 

\n ii) $dra_y$ is a quantifier of noncollinearity; this is actually clustering-independent (alias `democracy invariant' in Molecular Physics \cite{LR97}).
This is furthermore equal to $4 \m \times$ $area$ (the area of the triangle per unit $I$ in mass-weighted space), 
which is lucid enough to use as notation for this quantity. 
In comparison, in the equal-mass case 
\beq
\mbox{( physical area )} = \mbox{$\frac{I\sqrt{3}}{m}$} area \m  \label{area} \m . 
\eeq
\n iii) $dra_z$ is an ellipticity, $ellip$: 
the difference of the two `normalized' partial moments of inertia involved in the clustering in question, i.e.\ that of the base and that of the median.
This is clearly a function of pure ratio of relative separations, in contrast to $aniso$ being a function of pure relative angle.  
%
%FFFFFFFFFFFFFFFFFFFFFFFFFFFFFFFFFFFFFFFFFFFFFFFFFFFFFFFFFFFFFFFFFFFFFFFFFFFFFFFFFFFFFFFFFFFFFFFFFFFFFFFFFFFFFFFFFFFFFFFFFFFFFFFFFFFFFFFFFFFFFFFFFFFFFFFFFFFFFFFFFFFFFFFFFFFFFFFFFFFFFFFFF
{            \begin{figure}[!ht]
\centering
\includegraphics[width=1\textwidth]{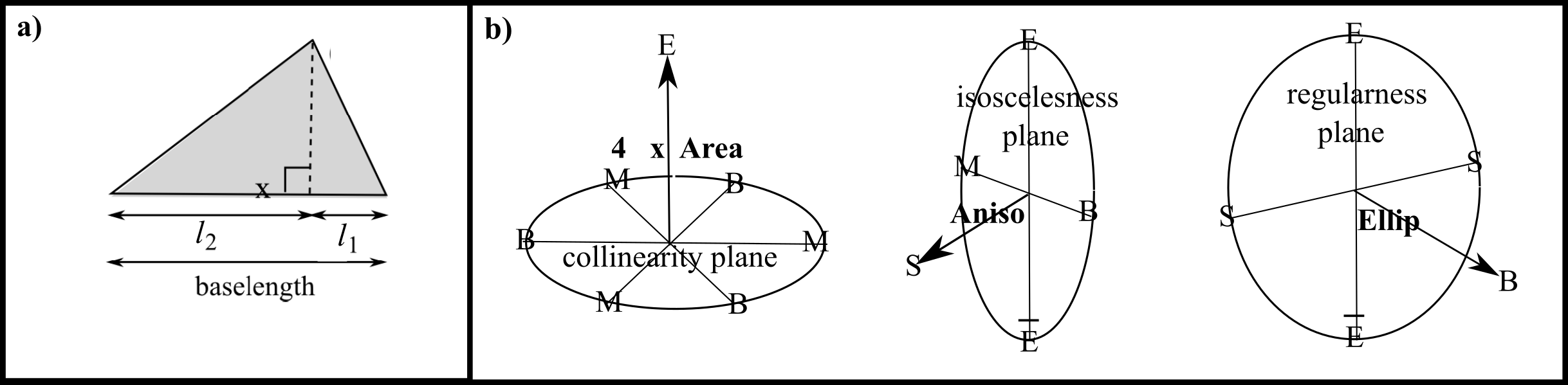}
\caption[Text der im Bilderverzeichnis auftaucht]{        \footnotesize{a) sets up the definition of anisoscelesness quantifier. 

\m 

\n 
b)-d) interpret the three `Hopf--Dragt' axes in terms of the physical significance of their perpendicular planes.} }
\label{Tutti-Tri-3} \end{figure}          }
%FFFFFFFFFFFFFFFFFFFFFFFFFFFFFFFFFFFFFFFFFFFFFFFFFFFFFFFFFFFFFFFFFFFFFFFFFFFFFFFFFFFFFFFFFFFFFFFFFFFFFFFFFFFFFFFFFFFFFFFFFFFFFFFFFFFFFFFFFFFFFFFFFFFFFFFFFFFFFFFFFFFFFFFFFFFFFFFFFFFFFFFFF

\m 

\n{\bf Remark 2} On the one hand, $\slTheta$ itself is also a ratio variable.  
Moreover, that $ellip$ is cos$\,\slTheta$ subsequently enters the mathematical study of triangleland (as the Legendre variable \cite{CH}).

\m 

\n On the other hand, $\slPhi$ is the relative angle `rightness variable' right corresponding to each clustering.
So, in contrast with the pure-ratio variable $dra_z = ellip$, $dra_x = aniso$ and $dra_y = 4 \times area$ provide mixed ratio and relative angle information. 

\m 

\n The ratio information in both of these of these is a $2\, n_1 n_2 = \sqrt{1 - ellip^2}$ factor. 

\m 

\n The relative angle information is in the cos$\,\slPhi$ and sin$\,\slPhi$ factors.   

\m 

\n{\bf Remark 3} See Fig \ref{Tutti-Tri-3}.b) for depiction of $Aniso$, $Ellip$ and $Area$ as vectors in relational space, 
identifying which plane therein each is perpendicular to and which shape space hemispheres this separates.

%===================================================================================================================================================================
\subsection{Outline of dynamical study of RPMs}
%===================================================================================================================================================================
%
%FFFFFFFFFFFFFFFFFFFFFFFFFFFFFFFFFFFFFFFFFFFFFFFFFFFFFFFFFFFFFFFFFFFFFFFFFFFFFFFFFFFFFFFFFFFFFFFFFFFFFFFFFFFFFFFFFFFFFFFFFFFFFFFFFFFFFFFFFFFFFFFFFFFFFFFFFFFFFFFFFFFFFFFFFFFFFFFFFFFFFFFFF
{            \begin{figure}[!ht]
\centering
\includegraphics[width=0.7\textwidth]{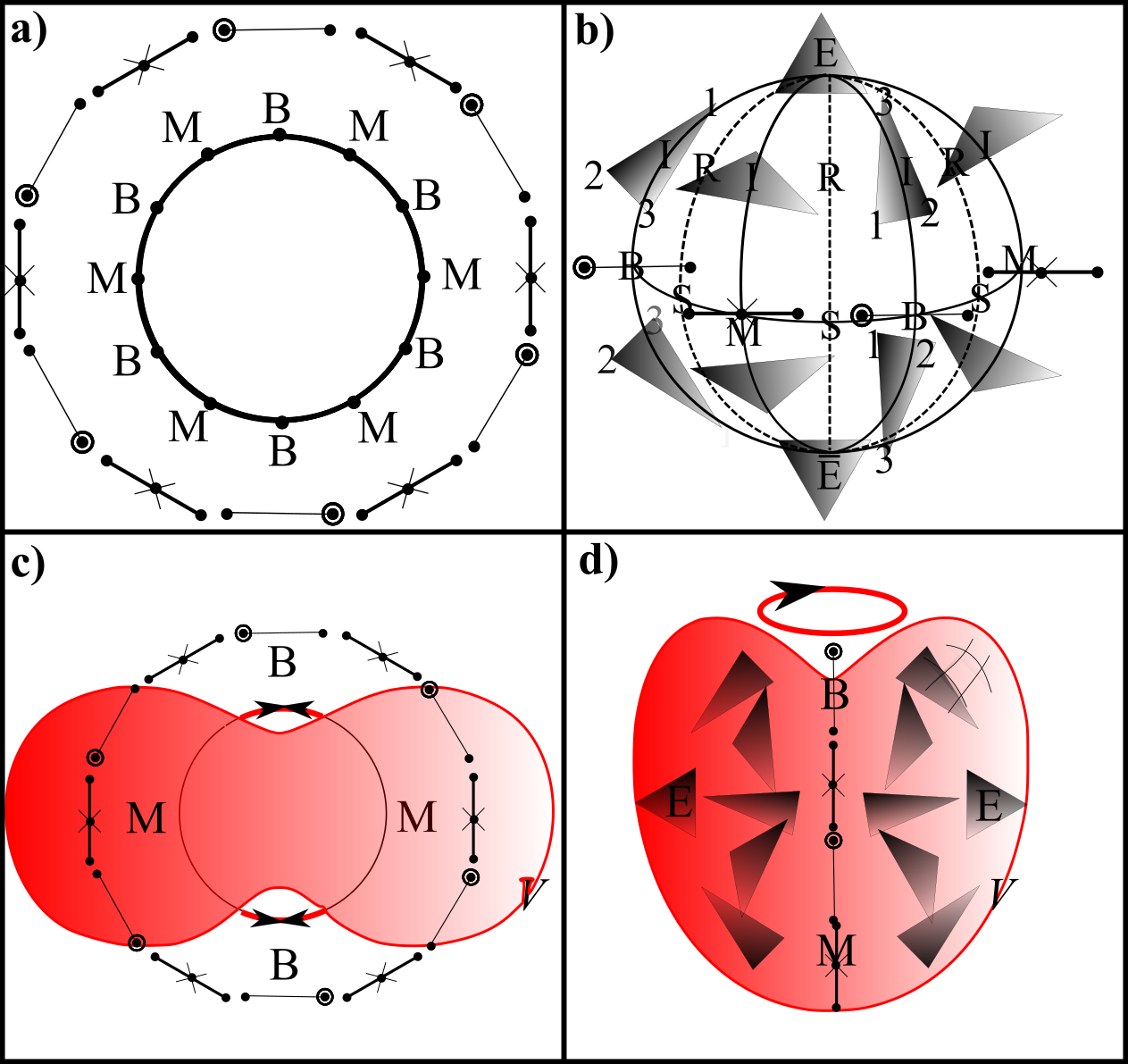}
\caption[Text der im Bilderverzeichnis auftaucht]{        \footnotesize{This figure is to be interpreted 
using Sec II.6's back-cloths and notation for configurations.

\m 

\n a) and b) are 3-stop metroland's and triangleland's free motion geodesics: round the loop and great circles respectively.  

\m 

\n c) and d) are examples of potentials over shape space -- indicated in red -- as induced by harmonic oscillator potentials: 
3-stop metroland's `peanut' and triangleland's `apple'. 
Note that d) is rotated by $\pi/2$ relative to b).  
The thick red lines denote some of the possible classical motions within the resulting potential wells.
The effective potentials in each case have an additional central skewer along the vertical axis, in the manner of a centrifugal barrier. 
These are relevant when the total shape momentum is nonzero.
See \cite{FileR} for more details and many more examples of potentials over -- and dynamical trajectories along -- shape spaces.} }
\label{Cl-RPM-Mosaic} \end{figure}          }
%FFFFFFFFFFFFFFFFFFFFFFFFFFFFFFFFFFFFFFFFFFFFFFFFFFFFFFFFFFFFFFFFFFFFFFFFFFFFFFFFFFFFFFFFFFFFFFFFFFFFFFFFFFFFFFFFFFFFFFFFFFFFFFFFFFFFFFFFFFFFFFFFFFFFFFFFFFFFFFFFFFFFFFFFFFFFFFFFFFFFFFFFF
 
\n{\bf Remark 1} Dynamical study of r-formulated RPMs proceeds as follows.
{\it Dilational momentum} $p_{\rho}$ is the momentum conjugate to the scale variable $\rho$, 
whereas the momenta conjugate to the shape variables $S^{\sfa}$ are, schematicly, the {\it shape momenta} $p^{\sS}_{\sfa}$.
For Metric Shape RPM in 1-$d$, $\FrQ$'s isometry group provides an $SO(n)$ of conserved quantities provided that the potential respects these.
These are relative dilational momentum quantities.
The 2-$d$ triangleland case has, under the same circumstances, an $SU(2)$'s worth of conserved quantities.
These are a pure relative angular momentum quantity (of the base relative to that of the median), 
and two mixtures of relative angular momentum and relative dilational momentum \cite{FileR}.
\cite{QuadI} gives the corresponding relational interpretation of quadrilateralland's $SU(3)$ octet of conserved quantities.  

\m

\n{\bf Remark 2} Fig \ref{Cl-RPM-Mosaic} subsequently presents some cases of potentials and dynamical trajectories for RPM's; 
these are useful as precursors and classical limits for quantum solutions used in Part III.
Free problem solutions are geodesics of the shape spaces.  
The physical significance of $\mathbb{S}^{2}$'s geodesic great circles in the cases of a) 4-stop metroland and b) triangleland.
    The $\mathbb{S}^{N - 2}$  geodesics -- corresponding to $N$-stop metroland free motions, 
and the $\mathbb{CP}^{N - 1}$ geodesics -- corresponding to $N$-a-gonland      free motions \cite{QuadI} -- are also well-known.      
The interpretation of the configurations can then just be read off the `back-cloth', such as in Figs 3 and 4.

%=========================================================================================================================================================================================
%=========================================================================================================================================================================================
\section{Examples of mass hierarchies and heavy--light ($h$-$l$) splits}
%=========================================================================================================================================================================================
%=========================================================================================================================================================================================

We finally expand our action in a useful precursor of various semiclassical approximations.

\m 

\n{\bf Approximation 1} [classical parallel of the Born--Oppenheimer-type split (IV.18)]. 
%
% Which originates in modelling electrons separately from the much heavier nuclei.
%
This is commonly used in both Classical and Quantum Cosmology \cite{KieferBook} 
as a means of modelling small inhomogeneities in a universe approximately modelled by a scalefactor and homogeneous matter terms.  

\m 

\n One consideration entering `heavy--light splits' is a mass ratio $m_l/m_h = \epsilon_{hl} \ll 1$.
This assumption is furthermore not made alone; e.g.\ `{\it sharply-peaked hierarchy}' conditions 
-- that all the $h$'s have similar masses $\gg$ all the similar masses of the $l$'s -- also enter at this stage: 
\be
\stackrel{\mbox{\scriptsize max}}  {  \mbox{\scriptsize $\ip$, $\jp$}   }\frac{|M_{\ip} -  M_{\jp}|}{M_{\ip}}      =:  \epsilon_{\Delta \sM} \ll 1 \mma
\stackrel{\mbox{\scriptsize max}}  {  \mbox{\scriptsize $\ipp$, $\jpp$} }\frac{|m_{\ipp} -  m_{\jpp}|}{m_{\ipp}}  =:  \epsilon_{\Delta \sm} \ll 1 \m .
\label{Sharp-Peak}
\ee
\be
\frac{m_{\ipp}  }{  M_{\ip}  }  \es 
\frac{    \frac{m_{\ipp} - m}{m}  m + m    }{    \frac{M_{\ip} - M}{M}M + M     } 
\m \m \sim \m \m \epsilon_{\sh\sll} \{ 1 + O( \epsilon_{\Delta \sM}, \epsilon_{\Delta \sm} )\}  \m . 
\ee
This allows for only one $h$-$l$ mass ratio to feature in subsequent approximations.  

\m

\n{\bf Example 1)} To meet the cosmological application, this Sec's particular $h$-$l$ split is aligned with the scale--shape split, 
whether of GR cosmology or of an RPM model arena of it.

\m 
 
\n{\bf Example 2)} Planck mass $m_{\sP\sll} \gg m_{\mbox{\scriptsize inflaton}}$ is a corresponding `gravitational mass hierarchy' 
which is sometimes used to motivate such approximations.  

\m 

\n{\bf Example 3)} In GR Cosmology, the scalefactor of the Universe dominates over one or both of the anisotropic or inhomogeneous modes. 
One could furthermore consider a two-step hierarchy which models both of these at once. 

\m

\n{\bf Remark 1} This classical treatment also begins to point to how some of the modelling assumptions made 
in the quantum cosmological version are in fact questionable.
For, these are qualitatively regime-dependent and so require justification rather than just being ushered in unchallenged.
This is a substantial insight since many such approximations look familiar due to having uses in other physical situations.
And yet the applicability of such approximations is regime-dependent 
and by no means guaranteed to carry over to the quantum cosmological regime for which they have now been proposed.  
Let us next pass to considering examples of $h$--$l$ splits.

\m
    
\n{\bf Example 4)} For Euclidean RPM, 1) the action is now 
\be
S  \es  \sqrt{2}\int \sqrt{E_{\sU\sn\si} - V_{h} - V_{l} - I} \sqrt{\d h^2 + h^2 ||\d \bil||_{\sbiM_{l}}\mbox{}^2   }
\m ,  
\label{JHL2}
\ee
(with $\u{B}$'s hung on the $\d l$'s in the indirectly formulated case) for 
$$
V_{h}  =  V_{h}( \mh        \mbox{ alone} ) 
       =  V_{\rho}(\rho)                     \m , \m 
V_{l}  =  V_{l}(l^{\sfa}    \mbox{ alone} ) 
       =  V_{\sS}(S^{\sfa}  \mbox{ alone} )  \m , \m
$$
\beq
I = I(h, l^{\sfa} \mbox{ alone})= I_{\rho\sS}(\rho, S^{\sfa}   \mbox{ alone}) \m . 
\ee
2) The conjugate momenta are (with $\Gamma = i I$ and a $\u{B}$ hung on each $\Last l$ in the uneliminated case)
\be
P^{h}   =  \Last h \m , \m P_{\Gamma}^{l} 
        =  h^2 M_{\Gamma\Lambda}\Last l^{\Lambda} \m .  
\label{HLmom2}
\ee
3) The classical energy constraint is 
\be
\scE  \:=  \half P_{h}^2 + \frac{||\biP_{l}||_{\sbiN_{l}}\mbox{}^2}{2 \, h^2 } + V_{h} + V_{l} + I  
      \es  E_{\sU\sn\si}                                                                                 \m .
\label{yet-another-E}
\ee
4) In the unreduced case, this is accompanied by  
\be
{\underline\scL_{l}}  \es  \sumfand \u{l}^{\Gamma} \cr \u{P}^{l}_{\Gamma}  \label{HLAMSha} \m .
\ee
5) The evolution equations are [in the same notation as eq (\ref{HLmom2})] 
\beq
\Last P^{h}           \es  h||\Last \bil||_{\sbiM_l}^2           \m - \m  \frac{\pa\{V_{h} + I\}}{\pa h}                                                   \mma
\Last P_{\Gamma}^{l}  \es  h^2 ||\Last \bil||^2_{\sbM_{l, \, \Gamma}}  \m - \m  \frac{\pa\{V_{l} + I\}}{\pa l^{\Gamma}}    \m .
\eeq
Here $\biM_l$ is the kinetic metric on $\bFrL$: the configuration space of the light degrees of freedom.  

\m 

\n 6) We can treat (\ref{yet-another-E}) in Lagrangian form as an equation for $t^{\se\sm}_{0}$ itself.
In the current classical setting, this is coupled to the $l$-equations of motion; moreover we need the $h$-equation to judge which terms to keep.
For more than one $h$ degree of freedom, these have separate physical content.    
The system is in general composed of the $E$-equation, $k_{h}$ -- 1 $h$-evolution equations and $k_{l}$ $l$-evolution equations.  

\m 

\n 7) The expression for the $t^{\se\sm}$ candidate is now (with the $\u{B}$'s and extremization thereover {\sl absent} in the eliminated case)
\be
\lt^{\se\sm}_{\sR\si}  \es  \mbox{\large E}^{\prime}_{\d \u{B} \, \in \,  Rot(d)}       
\int\sqrt{           \frac{   \d h^2 +  h^2||\d_{\u{B}} \bil||_{\sbiM_l}\mbox{}^2  }
                          {    2\{E_{\sU\sn\si} - V_{h} - V_{l} - I\}                      }            } \m .  
\label{TorreBruno}
\eeq
\n 1.h) The $h$-approximation to the action (\ref{JHL2}) is\footnote{$E_h$ is only approximately equal to $E_{\tU\tn\ti}$ 
%OOOOOOOOOOOOOOOOOOOOOOOOOOOOOOOOOOOOOOOOOOOOOOOOOOOOOOOOOOOOOOOOOOOOOOOOOOOOOOOOOOOOOOOOOOOOOOOOOOOOOOOOOOOOOOOOOOOOOOOOOOOOOOOOOOOOOOOOOOOOOOOOOOOOOOOOOOOOOOOOOOOOOOOOOOOOOOOOOOOOOOOOOO
since the $h$ and $l$ subsystems can exchange energy. 
$\Last^{h} := \pa/\pa t^{\te\tm(\tJ\tB\tB)}_{h}$.
Finally, 0 and 1 subscripts denote zeroth and first-order approximations.}  
%OOOOOOOOOOOOOOOOOOOOOOOOOOOOOOOOOOOOOOOOOOOOOOOOOOOOOOOOOOOOOOOOOOOOOOOOOOOOOOOOOOOOOOOOOOOOOOOOOOOOOOOOOOOOOOOOOOOOOOOOOOOOOOOOOOOOOOOOOOOOOOOOOOOOOOOOOOOOOOOOOOOOOOOOOOOOOOOOOOOOOOOOOO
\be 
S = \sqrt{2}\int \sqrt{\{E_{h} - V_{h}\}}\d h  \m . 
\ee 
2.h) The conjugate momenta are then $P^{h} = \Last^{h} h$, the quadratic energy constraint is 
\be 
\scE_{h} \:= \half \, P^{h\, 2} + V_{h}  
         \es  E_{h}                         \m . 
\ee  
3.h) As usual, the quadratic constraint can be taken as an equation for $t^{\se\sm}$ via the momentum--velocity relation.
The approximate emergent time candidate is 
\be
\lt^{\se\sm\mbox{-}0} \:= \lt^{\se\sm}_{h}  \:=  \int   \frac{ \d h_{0} }{ \sqrt{2\{E_{h} - V_{h_{0}}\}}  }  \m ,
\label{hint2} 
\ee
which is of the general form 
\beq
\lt^{\se\sm}_{h}  =  F[h, \d h]  \m . 
\label{hdh}
\eeq
0 here indicates zeroth order in $l$, i.e.\ the $h$-approximation.  

\m 

\n 4.h) For this split and to this level of approximation, there is no $\lFrg$-correction to carry out.
This is because the rotations act solely on the shapes and not on the scale.  
In other words Configurational Relationalism is trivial here.  

\m 

\n 5.h) The evolution equations are 
\be
\Last^{h} P^{h}   \es  -\frac{\pa V_{h}}{\pa h}  \m .
\ee
This assumes that (using the subscript j to denote `judging')  
$$
\mbox{(ratio of force terms)} \mma \mF_{\sj}  \:=  \frac{  \frac{\pa I}{\pa h}  }{  \frac{\pa V_{h}}{\pa h} }   
                                              \es  \frac{  \frac{\pa I}{\pa S}  }{ \frac{\pa V_{\tS}}{\pa S}   }  \m , 
$$
\beq
\m \m \mbox{is of magnitude } \m \epsilon_{\sss\sd\sss-1\sj} \ll  1 \m ,  
\label{SSA2}
\eeq
$$
\mbox{(ratio of geometrical terms)} \mma \mG_{\sj}  \:=    h    \frac{||\d \bil||^2_{\sbiM}}{\d^2 h} 
                                                    \es  \rho \frac{||\d \biS||^2_{\sbiM}}{\d^2 \rho}  \mma 
$$
\beq
\mbox{is of magnitude } \m \epsilon_{\sss\sd\sss-2\sj} \ll  1 \m .
\label{SSA3}
\eeq
The Author originally considered a `scale dominates shape' approximation \cite{FileR} at the level of the action, which is most clearly formulated as 
\beq
\mF  \:=  \frac{ I }{ W_h }  
     \es  \frac{ I }{ W_{\rho} } \mma \mbox{is of magnitude } \m \epsilon_{\sss\sd\sss-1} \ll 1 \m , 
\label{sds-1}
\eeq
\beq
\mG  \:=  \frac{||\d_{\u{B}}\bil||_{\sbiM}}{\d \, \mbox{ln} \, h }  
     \es  \frac{||\d_{\u{B}}\biS||_{\sbiM}}{\d \, \mbox{ln} \, \rho  } \mma \mbox{is of magnitude } \m \epsilon_{\sss\sd\sss-2} \ll 1            \m .
\eeq
Each pair -- i.e.\ 1, 1j, and 2, 2j -- are dimensionally the same but differ in further detail.
However, further consideration reveals that this assumption is better justified if judged at the level of the equations of motion and thus of forces. 

\m 

\n{\bf Remark 1} Whenever disagreement with experiment arises, 
going back to the previous Machian emergent time formulation should be perceived as a possible option.
Early 20th century `anomalous lunar motions' are an archetype for this \cite{DeSitter}.  
% 
% see also \cite{Bridgman27}+suppressout+ in this regard.  

\m

\n{\bf Remark 2} Moreover, pure-$h$ expressions of the general form (\ref{hdh}) are unsatisfactory 
from a Machian perspective since they do not give $l$-change an opportunity to contribute.  
This deficiency is to be resolved by treating them as zeroth-order approximations in an expansion involving the $l$-physics as well.
Expanding (\ref{TorreBruno}) in $l$, one obtains 
\beq
\lt^{\se\sm}_{\nFrg\mbox{\scriptsize -free \, 1}}  \es  \lt^{\se\sm}_{0} + \mbox{$\frac{1}{2\sqrt{2}}$}
\int \mbox{$\frac{d \rho}{W_{\rho}^{1/2}}
\left\{
\frac{I_{\rho S}}{W_{\rho}} +  
\left\{
\frac{\d S }{2 \, \d \, \mbox{\scriptsize ln} \,\rho}
\right\}^2 
\right\}$} 
+ O\left(\mbox{$\left\{\frac{I_{\rho S}}{W_{\rho}}\right\}^2 + \left\{\frac{\d S}{\d \, \mbox{\scriptsize ln} \,\rho}\right\}^4$}\right) \m , 
\label{Cl-Expansion}
\eeq
i.e.\ an expression of the form (compare \ref{hdh})
\beq
\lt^{\se\sm}_{\nFrg\mbox{\scriptsize -free \, 1}}  =  {\cal F}[h, l, \d h, \d l]  \m .  
\label{callie}
\eeq
More specifically, which moreover includes both an $l$-change term and an interaction term.  

\m

\n{\bf Remark 3} For comparison with the Semiclassical Approach, take note of classical adiabatic terms, which are related to the order of magnitude estimate    
\be
\epsilon_{\sA\sd}  \:=  \frac{\omega_{h}}{\omega_{l}} 
                   \es  \frac{t_{l}}{t_{h}}            \m ,
\ee
for $\omega_{h}$ and $\omega_{l}$ `characteristic frequencies' of the $h$ and $l$ subsystems respectively.   

\m 

\n{\bf Example 2)} One $h$--$l$ split in GR is scalefactor of th universe to other degrees of freedom.
Here
\be 
\mh_{ab} =: a^2\muu_{ab} \m , 
\ee 
so 
\be 
\{\d - \pounds_{\d{\usF}}\}\{a^2\muu_{ab}\}  \es  a^2\left\{\frac{\d a}{a} \muu_{ab} + \d \muu_{ab} - {\cal D}^{\su}\mbox{}_{(a}\d{\mF}_{b)} + 0   \right\}  
                                             \es  a^2\{\d - \pounds_{\d{\usF}}\}\muu_{ab}                                                                    \m ,
\ee											 
where the 0 arises from the constancy in space of the scalefactor-as-conformal-factor killing off the extra conformal connection. 
${\cal D}^{\su}_{a}$ here is the covariant derivative associated with the scale-free metric $\muu_{ab}$.]      
By this observation, scale--shape split approximate $t^{\se\sm}$ 
-- and its approximate semiclassical counterpart which coincides with it -- avoids the Best Matching Problem. 

\m 

\n In the present article's finite context, we can furthermore strike out the $\d F$ terms, corresponding to taking the minisuperspace subcase of GR. 

\m 

\n We next consider a classical Cosmology analogue of the astronomers' ephemeris time procedure; 
in Part III we extend this to Semiclassical Quantum Cosmology as well. 
Suppose that an accurate enough time has been found for one's purposes. 
One can then consider an analysis in terms of $(\biQ, \biP)$ as regards which features within that Universe contribute relevant change to the timestandard.  
This is very much expected to cover all uses of quantum perturbation theory that apply to modelling laboratory experiments.
Fairly large-scale features of the Universe are expected to contribute a small amount here, 
in addition to the zeroth-order expansion of the Universe and homogeneous matter mode contributions.  
There is a limit on such ephemeris time schemes, 
due to their iterations being at the level of form-fitting rather than a perturbative expansion of the equations of motion themselves.  

\m 

\n Let us point to minisuperspace models of anisotropy as one case of particular interest.  
For instance, diagonal Bianchi Class A models give rise to the following types of `scale dominates anisotropic shape' correction terms: 
\be 
\frac{V_{\mbox{\scriptsize$\bbeta$}}}{V_{\Omega}}    \=: \epsilon_{\sss\sd\sss-1} \m << \m  1
\ee 
(alongside a derivative version), and 
\be 
\frac{\d s_{\mbox{\scriptsize$\bbeta$}}}{\d \Omega}  \=: \epsilon_{\sss\sd\sss-2} \m << \m 1 \m .
\ee 
For Bianchi IX and VII$_0$, these are both $O(\mbox{anisotropy})^2$, whereas for Bianchi II, VI$_0$ and VIII the latter retains a $O(\mbox{anisotropy})$ piece.  
[For Bianchi I, there is no anisotropy potential, so the second type of correction term drops out altogether; see e.g.\cite{MacCallum} for the Bianchi classification.] 
These in turn induce as anisotropic ephemeris time corrections to cosmic time: 
\beq 
t_1^{\se\sm}  \es  t^{\scc\so\sss\sm\si\scc} + O(\mbox{anisotropy}^{1 \mbox{ \scriptsize or } 2}) \m .
\eeq  
\n See Article XI for more details of perturbative $h$--$l$ schemes, 
and Article XI for how pertubative treatment of inhomogeneity quite closely parallels the above treatment of anisotropy.

%=========================================================================================================================================================================================
\subsection{Problems with classical precursors of assumptions commonly made in Semiclassical Quantum Cosmology\label{Cl-Precursors}}
%=========================================================================================================================================================================================

{\bf Classical Problem 1)} Consider e.g.\ Newtonian Gravity or RPMs that model dust-filled GR cosmology. 
The corresponding regions of binary collision, B, the potential has infinite abysses and peaks (Fig \ref{Abyss}'s B lines are a simple example). 
`Scale dominates shape' approximations are thus certainly not valid near there, so some assumptions behind the Semiclassical Approach fail in the region around these lines.  
So for negative powers of relative separations, the heavy approximation only makes sense in certain wedges of angle. 
There is also the possibility that Dynamics set up to originally run in such regions falls out from them. 
These considerations point to a stability analysis being required to determine whether semiclassicality is representative. 
I.e.\ there is a tension between the procedure used in Semiclassical Quantum Cosmology 
and  the futility of {\sl trying to approximate a 3-body problem by a 2-body one} \cite{FileR}.  
%
%FFFFFFFFFFFFFFFFFFFFFFFFFFFFFFFFFFFFFFFFFFFFFFFFFFFFFFFFFFFFFFFFFFFFFFFFFFFFFFFFFFFFFFFFFFFFFFFFFFFFFFFFFFFFFFFFFFFFFFFFFFFFFFFFFFFFFFFFFFFFFFFFFFFFFFFFFFFFFFFFFFFFFFFFFFFFFFFFFFFFFFFFF
{            \begin{figure}[!ht]
\centering
\includegraphics[width=0.60\textwidth]{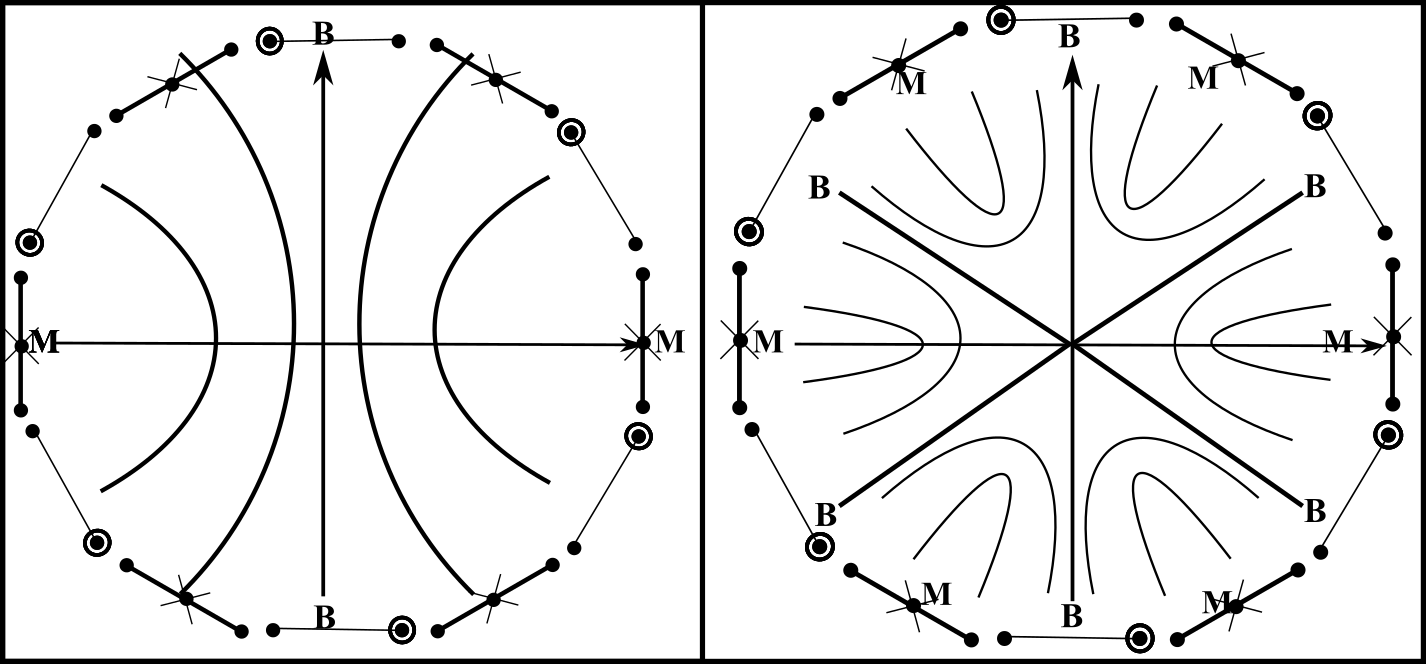}
\caption[Text der im Bilderverzeichnis auftaucht]{        \footnotesize{Contours on configuration space for single and triple negative power potentials 
(the 1-$d$ 3-particle case for simplicity). 
These have abysses along the corresponding binary collision lines B, to there being high ground in between these. 
(This is for negative-power coefficients such as for the attractive Newtonian Gravity potential.) 
M are merger configurations: the third particle is at the centre of mass of the other two. }        }
\label{Abyss}
\end{figure}  }
%FFFFFFFFFFFFFFFFFFFFFFFFFFFFFFFFFFFFFFFFFFFFFFFFFFFFFFFFFFFFFFFFFFFFFFFFFFFFFFFFFFFFFFFFFFFFFFFFFFFFFFFFFFFFFFFFFFFFFFFFFFFFFFFFFFFFFFFFFFFFFFFFFFFFFFFFFFFFFFFFFFFFFFFFFFFFFFFFFFFFFFFFF

\m 

\n{\bf Classical Problem 2)} Conventional treatments so far of Semiclassical Quantum Cosmology (Sec IV.3) decouple the $h$ and $l$ subsystems, 
which eases analytic solvability. 
This is partly attained through neglect of the $T_l$ term.  
However, the Classical Dynamics version of this (scale--shape split 1- or 2-$d$ RPM version) involves {\sl throwing away the central term}. 
I.e.\ the mathematical equivalent of neglecting the centrifugal barrier in the study of planetary motion. 
This causes unacceptable quantitative and qualitative errors (linear motion versus periodic motion along an ellipse).
Furthermore, this qualitative difference indeed carries over to the RPM counterpart \cite{FileR}.

%=====================================================BIBLIOGRAPHY==========================================================================================================================

\end{document}